\documentclass[a4paper,5p]{elsarticle}
\usepackage{amssymb}
\usepackage{amsmath}
\usepackage{graphicx}
\usepackage{xspace}
\usepackage{tabularx}
\usepackage{color}
\usepackage{cuted}

\usepackage{amsmath,amssymb,amsfonts,stmaryrd,wasysym,graphicx,multirow,textcomp,subfigure}
\usepackage{url}
\usepackage[colorlinks=true, citecolor=cyan, urlcolor=cyan]{hyperref}
\usepackage{romannum}
\usepackage{mathtools}
\usepackage[utf8]{inputenc}
\usepackage{braket}
\usepackage{xcolor}
\usepackage{color,soul} 
\usepackage{bm,xfrac}
\usepackage[normalem]{ulem}
\usepackage{enumitem}
\usepackage{dcolumn}
\usepackage{bm}
\usepackage[flushleft]{threeparttable}
\usepackage{array}
\usepackage{booktabs}
\usepackage{float}
\usepackage{setspace}
\usepackage{longtable}
\usepackage{balance}

\newcommand{\yk}[1]{{\color{black}{{#1}}}} 

\newcommand{\ve}{\varepsilon}

\usepackage{pdfpages} 
\usepackage{pgffor} 

\makeatletter
\AtBeginDocument{\let\LS@rot\@undefined}
\makeatother
\def\supplementfilename{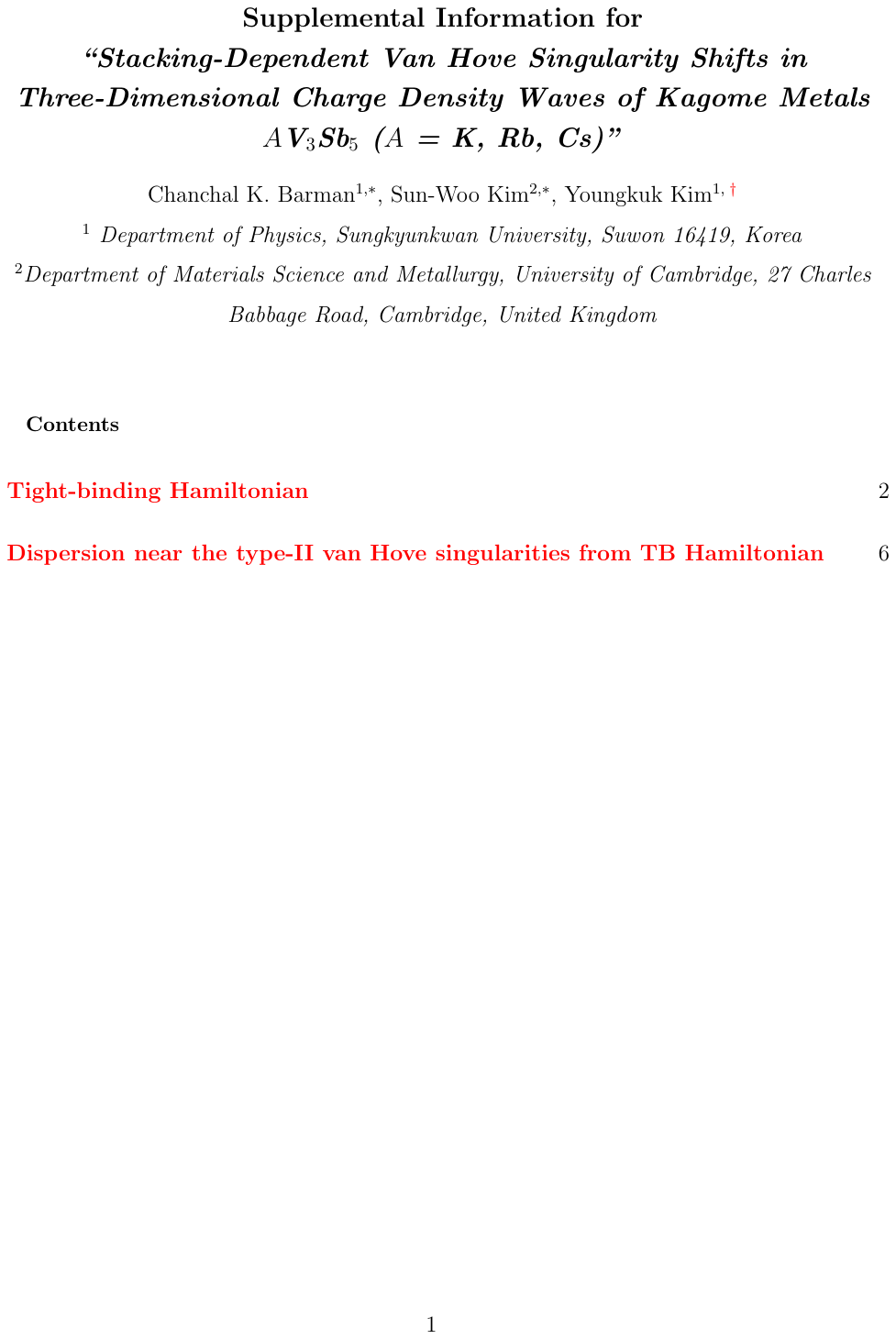}

\pdfximage{\supplementfilename}
\def\numbersupplementpages{\the\pdflastximagepages}

\newif\ifarXiv
\arXivtrue 

\begin{document}
\setcounter{page}{1}

\title{Stacking-Dependent Van Hove Singularity Shifts in Three-Dimensional Charge Density Waves of Kagome Metals $A$V$_3$Sb$_5$ ($A$ = K, Rb, Cs)}

\author[1]{Chanchal K. Barman\fnref{equal}}
\author[2]{Sun-Woo Kim\fnref{equal}}
\author[1]{Youngkuk Kim\corref{cor1}}
\ead{youngkuk@skku.edu}
\cortext[cor1]{Corresponding author}
\address[1]{Department of Physics, Sungkyunkwan University, Suwon 16419, Korea}
\address[2]{Department of Materials Science and Metallurgy, University of Cambridge, 27 Charles Babbage Road, Cambridge, United Kingdom}
\fntext[equal]{These authors contributed equally to this work.}
\date{\today}

\begin{abstract}
Vanadium-based kagomé systems $A$V$_3$Sb$_5$ ($A$ = K, Rb, Cs) have emerged as paradigmatic examples exhibiting unconventional charge density waves (CDWs) and superconductivity linked to van Hove singularities (VHSs). Despite extensive studies, the three-dimensional (3D) nature of CDW states in these systems remains elusive. This study employs first-principles density functional theory and a tight-binding model to investigate the stacking-dependent electronic structures of 3D CDWs in $A$V$_3$Sb$_5$, emphasizing the significant role of interlayer coupling in behaviors of the VHSs associated with diverse 3D CDW orders. We develop a minimal 3D tight-binding model and present a detailed analysis of band structures and density of states for various 3D CDW stacking configurations, including those with and without a $\pi$-phase shift stacking of the inverse star of David, as well as alternating stacking of the inverse star of David and the star of David. 
We find that VHSs exist below the Fermi level even in 3D CDWs without $\pi$-phase shift stackings, and that these VHSs shift downward in the $\pi$-phase shift stacking CDW structure, stabilizing the $2\times 2\times 2$ $\pi$-shifted inverse star of David distortions in alternating vanadium layers as the ground state 3D CDW order of $A$V$_3$Sb$_5$. Our work provides the electronic origin of 3D CDW orders, paving the way for a deeper understanding of CDWs and superconductivity in $A$V$_3$Sb$_5$ kagomé metals.
\end{abstract}

\maketitle

\section{Introduction} 
A kagomé lattice refers to a two-dimensional network of corner-sharing triangles. 
It offers a fertile ground for studying the intricate interplay between frustrated geometry, topology, and electronic correlations. 
The electronic energy band structure of a kagomé lattice generically features a flat band, VHSs, and a pair of Dirac points. 
Depending on the electron filling, various captivating quantum phenomena have been explored, such as quantum topology and geometry\,\cite{rhim2020quantum, li2020higher, kang2020topological}, superconductivity \cite{PhysRevB.79.214502,PhysRevB.85.144402,Kiesel_2013_uncoventional,Wang_2013_competing}, and quantum magnets \cite{yan2011spin, han2012fractionalized, yin2022topological, fu2015evidence, zhou2017quantum}. The kagomé lattice in stable materials has been a target of rigorous research efforts, resulting in many experimentally verified materials.
Notable examples include FeSn\,\cite{kang2020dirac}, Fe$_3$Sn$_2$\,\cite{lin2018flatbands}, and CoSn\,\cite{huang2022flat}, which illustrates physics originating from flatbands. 
More recently, kagomé metals such as $A$V$_3$Sb$_5$ \cite{ortiz2019new,Oritz_2020_SC1,Wilson2024} and $A
$Ti$_3$Bi$_5$\,\cite{yang2022titanium,WerhahnOrtizHayWilsonSeshadriJohrendt,liu2023tunable} for $A$ = K, Rb, or Cs and $R$V$_6$Sn$_6$\,\cite{pokharel_electronic_2021,arachchige_charge_2022,wang2024origin} ($R$ = rare earth), and FeGe\,\cite{Teng2022_Fege1,Fege-2,Fege-3}, have been under intensive investigation to study phenomena associated with the VHSs. 

The vanadium-based kagomé metal $A$V$_3$Sb$_5$ ($A$ = Cs, K, and Rb) has become a prototypical system within the family of kagomé metals. It hosts multiple VHSs near Fermi level that exhibit rich properties\,\cite{Wu_2021_Nature, kang2022twofold, Hu2022_rich, Kim_2023_ML_AVS}, and these VHSs have been found to play a significant role in driving unconventional phenomena\,\yk{\cite{Neupert2022, Jiang_2022_kagome}}. 
Notably, the onset of an anomalous Hall effect\,\cite{Yang_unconventional_AHE1,Yu_2021_AHE2} occurs concurrently with the CDW phase transition below $T_{\text{CDW}}\approx80-100$\, K\,\cite{Oritz_2020_SC1, Oritz_2021_SC2, Yin2021_CDW, Jiang_2021_Nat_Mat_STM}.
The presence of Hall conductivity signals the time-reversal symmetry breaking\,\cite{LC_theory_1,LC_theory_2,LC_theory_3,LC_theory_4}, which is further evidenced by various experiment methods\,\cite{TRSB_expt_1,TRSB_expt_2,TRSB_expt_3}.
Moreover, six-fold rotation symmetry breaking at $T_{\text{CDW}}$ is observed\,\cite{Xiang2021_six_breaking_1,Li2022_six_breaking_2,Luo2022_six_breaking_3,Li2023_six_breaking_4,Wu_2022_six_breaking_5,Wulferding_2022_six_breaking_6,kang2023charge_six_breaking_7,Wu_2022_six_breaking_8,Jiang2023_six_breaking_9}, followed by additional rotational symmetry breaking at lower temperatures\,\cite{Nie2022_nematic_1,Sur2023_nematic_2}, which is assigned as electronic nematicity \,\cite{TRSB_expt_3, Jiang2023_six_breaking_9,Nie2022_nematic_1}.
Upon further cooling, superconductivity emerges with a critical temperature $T$$_{\mathrm{c}}$ $\sim$ 0.9–2.5\,K\,\cite{Oritz_2020_SC1, Oritz_2021_SC2, Yin2021_CDW} and
the superconductivity is found to compete with CDW\,\cite{kang2023charge_six_breaking_7,chen2021double,Yu2021,Zheng2022}

\begin{table*}[t!]
\centering
\caption{TB parameters for $A$V$_3$Sb$_5$ (in eV). $\ve$ is the onsite energy. $t$ and $t_2$ are the nearest-neighbor (NN) and next-nearest-neighbor (NNN) hopping parameters in the kagomé plane [Fig.\,\ref{fig:structure}(c)]. $t_s$ is the strength of the CDW order parameter [Figs.\,\ref{fig:structure}(d) and\,\ref{fig:structure}(e)]. $t_{z1}$, $t_{z2}$, $t_{z3}$, and $t_{z4}$ are NN, NNN, NNNN, and NNNNN hopping strengths along the out-of-plane direction [Fig.\,\ref{fig:structure}(f)].}
\label{table:TBparam}
\begin{tabular}{ccccccccc}
\hline
    & $\ve$ & $t$ & $t_{2}$ & $t_{s}$ & $t_{z1}$ & $t_{z2}$ & $t_{z3}$ & $t_{z4}$\\ \hline
KV$_3$Sb$_5$  & -0.010 & 0.330 & 0.045 & 0.040 & 0.048 & -0.009 & 0.018 & -0.006  \\
RbV$_3$Sb$_5$ & 0.020 & 0.340 & 0.036 & 0.050 & 0.048 & -0.007 & 0.022 & -0.004 \\
CsV$_3$Sb$_5$ & 0.080 & 0.365 & 0.010 & 0.060 & 0.048 & -0.006 & 0.030 & -0.004 \\
CsV$_3$Sb$_5$ (SD+ISD) & 0.100 & 0.380 & 0.025 & 0.060 & 0.048 & -0.006 & 0.030 & -0.004 \\
\hline
\end{tabular}
\end{table*}

Despite extensive studies\,\cite{LC_theory_4,tan2021charge,li2021observation,Ptok22p235134,PhysRevB.105.045135,liu_observation_2022,he_anharmonic_2024,gutierrez2023phonon}, the underlying mechanism of the CDW state, particularly its 3D nature, is not fully understood.
It has been well established that the in-plane CDW pattern is either $2\times 2$ star of David (SD) or inverse star of David (ISD) type distortion in the single vanadium kagomé layer\,\cite{kang2022twofold,tan2021charge,ortiz2021fermi}.
The CDW state in $A$V$_3$Sb$_5$ exhibits a 3D nature with a modulation along the $c$ (out-of-plane) axis. Moreover, CsV$_3$Sb$_5$ is also found to exhibit a CDW state with a $2\times 2\times 4$ periodicity under certain conditions\,\cite{ortiz2021fermi,stahl2022temperature,xiao2023coexistence,kautzsch2023structural}, distinguishing it from KV$_3$Sb$_5$ and RbV$_3$Sb$_5$. \yk{The energetics of VHSs responsible for the emergence of 3D CDW can provide insight into these phenomena, as the presence of VHSs near the Fermi level is known to contribute to various phenomena, such as superconductivity \cite{markiewicz971179, Kim18p165102, Wan23p41}, magnetism \cite{Irkhin01p165107, Mazin97p2556}, and nematicity \cite{Yamase05p035114}.} Nevertheless, the origin of such intricate CDW order in $A$V$_3$Sb$_5$ remains a topic of ongoing research.

In this paper, we use first-principles density functional theory (DFT) and a tight-binding (TB) model to investigate the influence of interlayer interactions on the electronic structure of $A$V$_3$Sb$_5$ ($A$ = K, Rb, Cs) kagomé systems, with a particular emphasis on changes in the VHS near the Fermi level depending on various 3D CDW stacking configurations. We develop a minimal 3D TB model that effectively captures interlayer interactions. This model faithfully reproduces the DFT bands and density of states for all three $A$V$_3$Sb$_5$ compounds.
Our investigation reveals the critical role of interlayer interactions among vanadium layers in describing the diverging electronic density of states near the Fermi level, which influences the ground state 3D CDW order. 
Based on our comprehensive calculations, we demonstrate that in 3D CDW structures without a $\pi$-phase shift, the VHS is close to the Fermi level. However, in the ISD structure with lateral $\pi$-phase shift stacking, the VHS moves sufficiently below the Fermi level, stabilizing this configuration as the ground state 3D CDW order.
Our findings contribute to the understanding of the origin of 3D CDW orders observed in kagomé metals.

\section{Methods}
\subsection{DFT methods}
To investigate the electronic structures of $A$V$_3$Sb$_5$ ($A$ = K, Rb, SC), we performed first-principles calculations based on density functional theory (DFT) as implemented in the Vienna {\em ab initio} simulation package (VASP)\,\cite{PhysRevB.47.558, VASP1, VASP2, Kresse99p1758}. We used Perdew-Burke-Ernzerhof (PBE)\,\cite{perdew1996generalized} within the generalized gradient approximation (GGA) to describe the exchange and correlation functionals. We used projector augmented wave (PAW) pseudopotentials \cite{blochl1994projector} and plane basis set with an energy cutoff of 300 eV. We applied the zero-damping DFT-D3\,\cite{grimme2010consistent} van der Waals correction throughout our calculations. Brillouin zone integration was performed using 
the $\Gamma$-centered sampling with $11\times 11 \times 13$ $k$-mesh for the $2\times2\times1$ unit cell and $11\times 11 \times 7$ $k$-mesh for the $2\times2\times2$ unit cell.
The geometry of the structures was optimised until all forces are below 0.01\,eV/\AA.
We used the Wannier90 package\,\cite{Pizzi_2020} to generate maximally localized Wannier functions (MLWFs)\,\cite{PhysRevB.56.12847,PhysRevB.65.035109,RevModPhys.84.1419}, enabling us to obtain the Fermi surface (FS). Subsequently, we extracted constant energy contours from the FS using the FermiSurfer tool\,\cite{fermisurfer}.

\begin{figure}[t!]
\centering
\includegraphics[width=.5\textwidth]{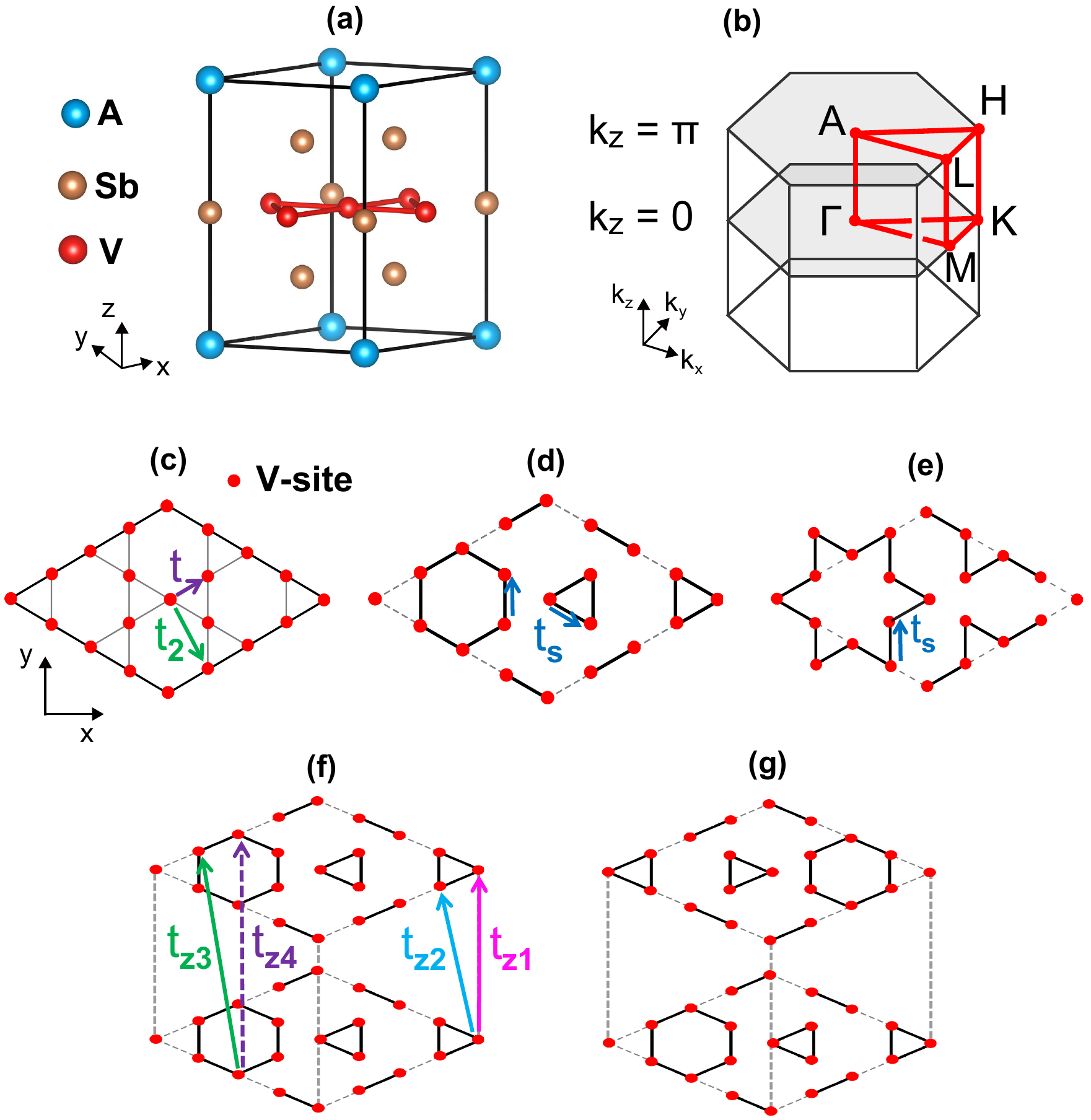}
\caption{(a) Primitive $1 \times 1 \times 1$ unit cell of $A$V$_3$Sb$_5$ ($A$ = K, Rb, Cs). (b) Hexagonal Brillouin zone. The $k_z = 0$ and $k_z = \pi$ planes are indicated by grey planes. High-symmetry lines and points are colored red. (c) Top view of the vanadium kagomé layer in the $2 \times 2 \times 1$ cell. Red circles represent the vanadium sites. (d) Inverse star of David and (e) Star of David CDW structures in the in-plane kagomé layer. The connected solid lines represent the shortened bonding lengths compared to the pristine structure. (f,g) $2 \times 2 \times 2$ supercell (f) without $\pi$ phase shift and (g) with $\pi$ phase shift stackings of ISD layers along the out-of-plane direction. Arrows schematically delineate the hopping of electrons via the corresponding hopping parameters $t$, $t_2$, $t_s$, $t_{z1}$, $t_{z2}$, $t_{z3}$, and $t_{z4}$, respectively.}
\label{fig:structure}
\end{figure}

\begin{figure*}[t!]
\centering
\includegraphics[width=.9\textwidth]{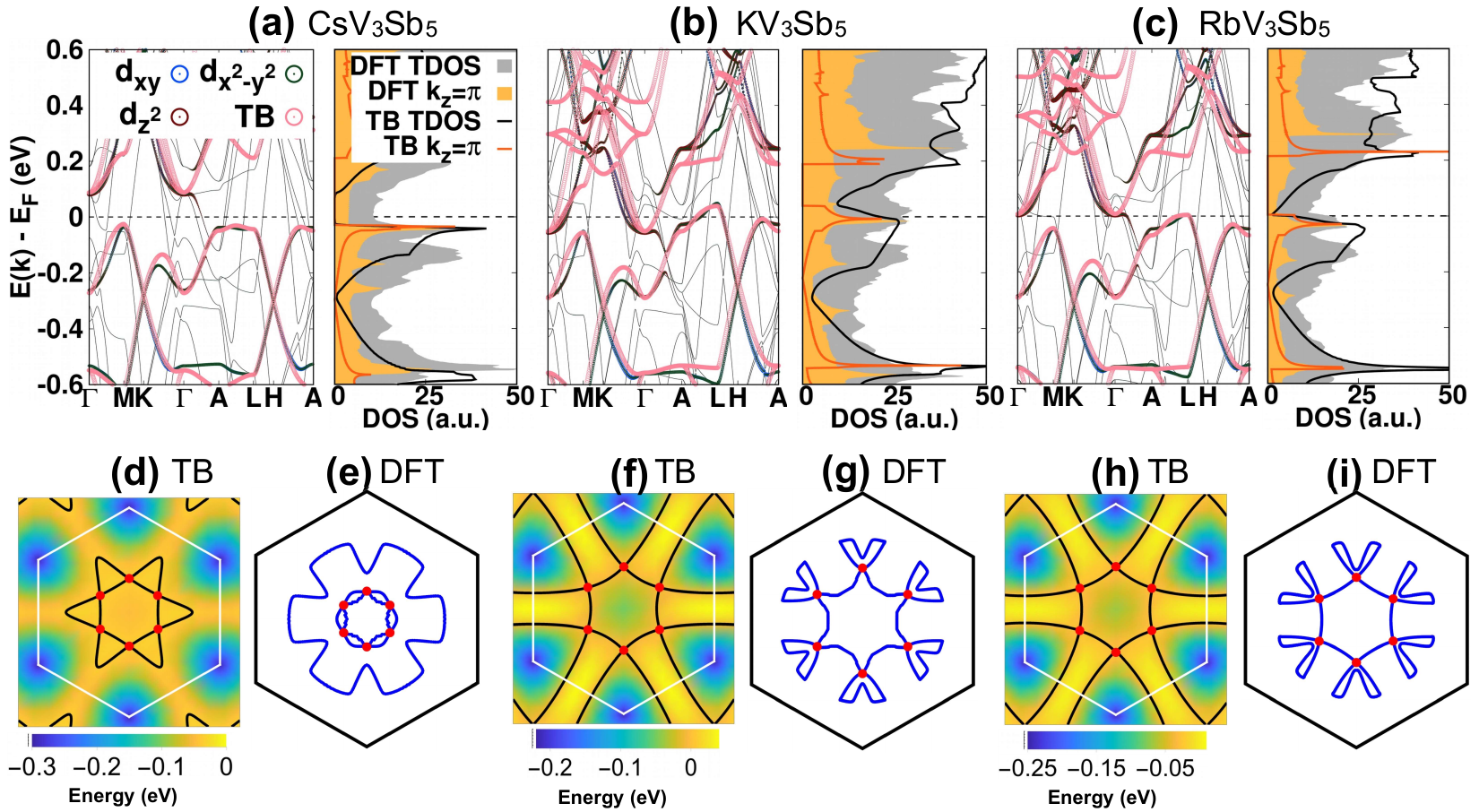}
\caption{(a-c) DFT and TB electronic band structures and density of states (DOS) of the $2 \times 2 \times 1$ ISD structures for (a) CsV$_3$Sb$_5$, (b) KV$_3$Sb$_5$, and (c) RbV$_3$Sb$_5$. 
In these plots, the DFT band structure is projected onto the $d_{xy}$, $d_{3z^2-r^2}$, and $d_{x^2-y^2}$ orbitals (in global coordinates) of vanadium atoms, with the magnitude of open circles proportional to the projected weight.
For the DOS, we display both the total DOS (TDOS) and the DOS resolved for the $k_z=\pi$ plane.
(d,e) Constant energy contours in the $k_z=\pi$ plane for CsV$_3$Sb$_5$ obtained from (d) the TB and (e) the DFT calculations. (f,g) and (h,i) show the same as (d,e) but for KV$_3$Sb$_5$ and RbV$_3$Sb$_5$, respectively. Energy contours using TB (DFT) are taken at -0.0346 (-0.0364) eV, -0.0070 (0.0097) eV, and -0.0287 (-0.0336) eV for CsV$_3$Sb$_5$, KV$_3$Sb$_5$, and RbV$_3$Sb$_5$, respectively. Red colored dots designate the VHS points in the energy contours.}
\label{fig:ISDband}
\end{figure*}

\subsection{Tight-binding model}


We construct a 24-band tight-binding model that describes the $d_{xy}$ orbitals (in local coordinates) of the vanadium kagomé layers within a $2\times 2\times 2$ supercell.
The $d_{xy}$-orbital tight-binding model gives rise to the $A_g$ irreducible representation for the VHS state at the M point near the Fermi level in the $1 \times 1 \times 1$ pristine structure (see also Ref.\,\cite{Kim_2023_ML_AVS}).
The amplitude of the hopping parameters was determined so that the DFT bands were reproduced. We introduce a twelve-component spinor in the $2 \times 2 \times 1$ unit cell of the vanadium kagomé layer and write the tight-binding Hamiltonian as
\[
H = \sum_{\boldsymbol{k}} \Psi^{\dagger}_{\boldsymbol{k}} \mathcal{H}(\boldsymbol{k}) \Psi_{\boldsymbol{k}}.
\]
The spinor \(\Psi_{\boldsymbol{k}}\) is composed of three different sublattices \(A_{\boldsymbol{k}}, B_{\boldsymbol{k}},\) and \(C_{\boldsymbol{k}}\), where
\[
\alpha_{\boldsymbol{k}} = (\alpha_{1\boldsymbol{k}}, \alpha_{2 \boldsymbol{k}}, \alpha_{3\boldsymbol{k}}, \alpha_{4\boldsymbol{k}})^{\rm T}
\]
for \(\alpha = A, B, C\). The \(\alpha_i\) refers to the \(\alpha\) sublattice in the \(i\)-th site within the \(2 \times 2 \times 1\) supercell, as shown in Fig.\,\ref{fig:structure} (see also Fig.\,S1 in the Supplemental Information).
Our minimal TB Hamiltonian can be grouped into
\begin{align}
\mathcal{H}(\boldsymbol{k}) &= \mathcal{H}_{\rm NN}(\boldsymbol{k}) + \mathcal{H}_{\rm NNN}(\boldsymbol{k}) \\
 & + \mathcal{H}_{\rm SD}(\boldsymbol{k}) + \mathcal{H}_{\rm ISD}(\boldsymbol{k}) + \mathcal{H}^{\rm out}(\boldsymbol{k}) \notag
\end{align}
Here, $\mathcal{H}_{\rm NN}(\boldsymbol{k})$ and $\mathcal{H}_{\rm NNN}(\boldsymbol{k})$ are the nearest-neighbor (NN) and next-nearest-neighbor (NNN) hopping matrices within the 2D kagomé layer. 
$\mathcal{H}_{\rm SD}(\boldsymbol{k})$ and $\mathcal{H}_{\rm ISD}(\boldsymbol{k})$ are the NN hopping matrices that describe the SD and ISD type distortions [Figs.\,\ref{fig:structure}(d) and\,\ref{fig:structure}(e)], respectively. $\mathcal{H}^{\rm out}(\boldsymbol{k})$ 
accounts for the interlayer hoppings between out-of-plane vanadium sites [Fig.\,\ref{fig:structure}(f)], specified by the level of nearest neighbors as
\begin{align}
\mathcal{H}^{\rm out}(\boldsymbol{k}) &= \mathcal{H}_{\rm NN}^{\rm 3D}(\boldsymbol{k}) + \mathcal{H}_{\rm NNN}^{\rm 3D}(\boldsymbol{k}) \\
 & + \mathcal{H}_{\rm NNNN}^{\rm 3D}(\boldsymbol{k}) + \mathcal{H}_{\rm NNNNN}^{\rm 3D}(\boldsymbol{k}) \notag
\end{align}
The detailed forms of these matrices are given in the Supplemental Information\footnote{See the detailed forms of matrices of the tight-binding Hamiltonian}.
\yk{Previous tight-binding models have been developed to accurately describe multiple VHSs near the Fermi energy \cite{Jeong22p235145, li2023origin} and band topology \cite{deng2023fermi}, including the Sb \(p\) orbital. Our model differs from these models in that it excludes the Sb \(p\) orbital and focuses on generically describing a single VHS near the Fermi level originating from the \(k_z=\pi\) plane that is sensitive to the 3D CDW formation. This approach focuses on the instability in the $2\times2\times1$ supercell structure to explore the instability in the 3D stacking configurations of the CDW phases.
}

\section{Results and Discussion}
Let us first briefly introduce the atomic structure of $A$V$_3$Sb$_5$ in the pristine and CDW phases. The pristine phase of $A$V$_3$Sb$_5$ has a layered hexagonal structure (space group $P6/mmm$, No.\,191). This structure features specific atomic positions and stacking configurations, shown in Fig.\,\ref{fig:structure}(a). The corresponding hexagonal Brillouin zone is depicted in Fig.\,\ref{fig:structure}(b). In the pristine structure, the alkali atoms $A$ occupy the Wyckoff position $1a(0, 0, 0)$. The kagomé layers consist of vanadium atoms located at the Wyckoff position $3g(\frac{1}{2},\frac{1}{2},\frac{1}{2})$, interleaved with a hexagonal lattice of Sb atoms situated at the Wyckoff position $1b (0,0,\frac{1}{2})$. Another layer of Sb atoms with a honeycomb structure resides at Wyckoff position $4h(\frac{2}{3}, \frac{1}{3}, z_{Sb})$. The atomic parameters slightly vary depending on the $A$ atom. The optimized structural parameters, in good agreement with the experimental parameters, are listed in Table\,\ref{table:structureParams}.

\begin{table}[b!]
\centering
\caption{Structural parameters for pristine $A$V$_3$Sb$_5$ ($A$ = Cs, K, Rb) obtained from DFT. The lattice parameters ($a,c$) are given in the Angstrom unit, while $z_{Sb}$ is provided in the fractional unit of $c$. Experimental values from Ref.\,\cite{ortiz2019new} are given in parenthesis for comparison. }
\label{table:structureParams}
\begin{tabular}{cccc}
 $A$   & $a$\,[\AA] & $c$\,[\AA] & $z_{Sb}$\,[c]\\ \hline
Cs   & 5.439 (5.495) &    9.326 (9.309) & 0.742 (0.742) \\
K    & 5.411 (5.482) &    8.893 (8.948) & 0.756 (0.754) \\
Rb   & 5.425 (5.472) &    9.114 (9.073) & 0.749 (0.750)\\
\hline
\end{tabular}
\end{table}

\begin{figure*}[t!]
\centering
\includegraphics[width=.65\textwidth]{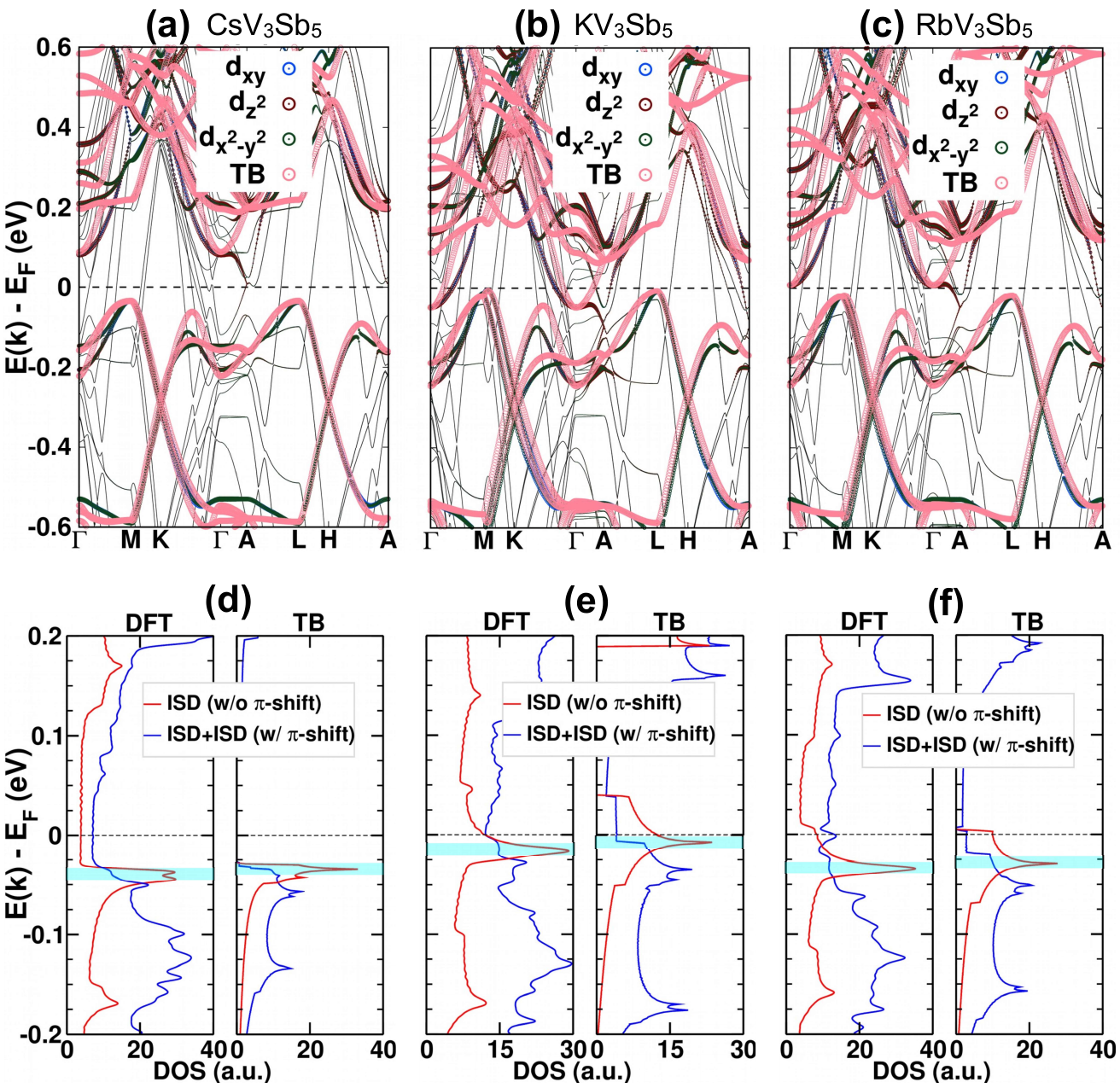}
\caption{
(a-c) DFT and TB electronic band structures and DOS of the $2 \times 2 \times 2$ $\pi$-phase shift ISD+ISD structures for (a) CsV$_3$Sb$_5$, (b) KV$_3$Sb$_5$, and (c) RbV$_3$Sb$_5$. 
The orbital-projected DFT band structures, with the radius of open circles proportional to the projected weight, are overlaid with the TB band structures.  
(d-f) The $k_z=0$ (and the $k_z=\pi$) plane-resolved DOS of the $2 \times 2 \times 2$ $\pi$-shifted ISD+ISD (and the $2 \times 2 \times 1$ non-$\pi$-shifted ISD) structures obtained using DFT and TB for (d) CsV$_3$Sb$_5$, (e) KV$_3$Sb$_5$, and (f) RbV$_3$Sb$_5$. \yk{The DFT DOSs are normalized per unit cell. We note that ISD+ISD has double the number of atoms compared to the ISD.}
}
\label{fig:ISDISD}
\end{figure*}

For the CDW structures, we consider both ISD and SD type distortions observed in the experiment\,\cite{ortiz2021fermi}.
Figures\,\ref{fig:structure}(c-e) show top views of the vanadium kagomé layer for the pristine, the ISD, and SD  structures, respectively. 
Compared to the $1\times 1$ pristine structure, the ISD and SD distortions expand to a $2\times 2$ in-plane periodicity. 
The stacking configurations of non-$\pi$-phase shifted and $\pi$-phase shifted ISD+ISD type kagomé layers along the out-of-plane $c$-direction are illustrated in Figs.\,\ref{fig:structure}(f) and \ref{fig:structure}(g). 
The non-$\pi$-shifted stacking configuration has a lattice periodicity of $2\times 2\times 1$ with a uniform atomic arrangement without any in-plane translation between the adjacent kagomé layers.
Conversely, the $\pi$-shifted configuration involves a half translation of the in-plane unit vectors between adjacent kagomé layers, leading to a $2\times 2\times 2$ periodicity. In line with the previous calculations \cite{tan2021charge,subedi2022hexagonal},  our DFT total energy calculations show that the $\pi$-shifted $2\times 2\times 2$ ISD+ISD phase is energetically more favorable than the non-$\pi$-shifted phase for all three alkali atoms ($A$ = K, Rb, Cs). The energy decreases by 16.8, 16.6, and 10.6 meV per $2\times2\times 1$ unit cell for the $\pi$-shifted phase compared to the non-$\pi$-shifted phase in KV$_3$Sb$_5$, RbV$_3$Sb$_5$, and CsV$_3$Sb$_5$, respectively.

Our TB model can generically reproduce the DFT bands for various CDW stacking configurations. A specific set of TB parameters accurately replicates the DFT bands for the $d_{xy}$, $d_{3z^2-r^2}$, and $d_{x^2-y^2}$ orbitals (in global coordinates) near the Fermi level for both the $\pi$-shifted and non-$\pi$-shifted stacking configurations of ISD or SD structures. Figures~\ref{fig:ISDband}(a-c) depict the electronic band structures and the density of states obtained using both DFT and TB for the $2\times2\times 1$ non-$\pi$-shifted ISD structure for all three alkali metals. Our TB bands (colored pink) closely match the DFT bands of vanadium $d_{xy}$, $d_{3z^2-r^2}$, and $d_{x^2-y^2}$ orbitals (colored blue, brown, and green, respectively) for all three alkali metals. The employed sets of TB parameters for $A$V$_3$Sb$_5$ ($A$ = K, Rb, and Cs) are presented in Table\,\ref{table:TBparam}. 
The out-of-plane hopping parameters in our minimal vanadium TB models show substantial magnitudes, which effectively capture the three-dimensional nature of these systems. Notably, the nearest-neighbor (NN) hopping parameter along the out-of-plane direction, $t_{z1}$, is larger than the in-plane next-nearest-neighbor (NNN) hopping parameter, $t_2$, for all three systems. Additionally, the NNNN out-of-plane hopping parameter $t_{z3}$ is larger than the NNN out-of-plane hopping parameter $t_{z2}$, which we attribute to the weak overlap between the NNN sites.

\begin{table}[t!]
\centering
\caption{Relative total energy (in meV per $2\times2\times 1$ unit cell) for various CDW stacking configurations. The total energy of the non-$\pi$-shifted ISD stacking is set to zero.}
\label{table:sd+isd}
\begin{tabular}{cccc}
&  non-$\pi$-shift & non-$\pi$-shift  & $\pi$-shift  \\ 
&  ISD+ISD & SD+ISD  & ISD+ISD \\ \hline
 CsV$_3$Sb$_5$ & 0.0 & 15.4  & -10.6 \\
 \hline
\end{tabular}
\end{table}

Our comprehensive DFT and TB calculations reveal a sharp peak in the DOS just below the Fermi level for the $2\times2\times 1$ non-$\pi$-shifted ISD structure in all three compounds, mainly attributed to the VHS at $k_z = \pi$. 
The constant energy contours in the $k_z = \pi$ plane, obtained by both DFT [Figs.\,\ref{fig:ISDband}(e,g,i)] and TB [Figs.\,\ref{fig:ISDband}(d,f,h)] calculations, show six symmetry-related van Hove saddle points, with one located along the $A-H$ direction and the other five along other symmetry-related high symmetry lines.
The VHSs are identified by the crossing points (colored red) in the constant energy contour, which satisfy two conditions: (i) zero gradient of energy dispersion $\nabla_{\mathbf {k}}E_{\mathbf {k}}=0$ and (ii) a negative value of the Hessian determinant $ \frac{\partial^2 E_{\mathbf {k}}}{\partial k_x^2} \frac{\partial^2 E_{\mathbf {k}}}{\partial k_y^2} - \frac{\partial^2 E_{\mathbf {k}}}{\partial k_x \partial k_y} \frac{\partial^2 E_{\mathbf {k}}}{\partial k_y \partial k_x} < 0$.
These VHSs at the $k_z=\pi$ plane primarily contribute to the large DOS below the Fermi level, as consistently confirmed by both DFT and TB calculations [Figs.\,\ref{fig:ISDband}(a-c)], indicating that the $d_{xy}$, $d_{3z^2-r^2}$, and $d_{x^2-y^2}$ orbitals of vanadium atoms (in global coordinates) are responsible for electronic instabilities near the Fermi level in the non-\yk{$\pi$}-shifted CDW order.

We conduct further analysis on the VHSs in the $2\times2\times 1$ non-$\pi$-shifted ISD structure.
The VHSs are classified as type-II, as they reside off time-reversal invariant momenta\,\cite{Kim_2023_ML_AVS,PhysRevB.92.035132,Yuan2019}.
The explicitly calculated Hessian determinant values using TB models for these type-II VHSs are -0.327 eV\AA$^2$, -1.716 eV\AA$^2$, and -1.408 eV\AA$^2$ for CsV$_3$Sb$_5$, KV$_3$Sb$_5$, and RbV$_3$Sb$_5$, respectively.
The value is quite close to zero in the case of Cs, in line with the flat-like dispersion along the $A-L$ direction [Fig.\,\ref{fig:ISDband}(a)] and quasi-flat energy band landscape across broader regions near the A point [Fig.\,\ref{fig:ISDband}(d)], a distinct feature from the other two systems.
Consequently, the VHSs in the $2\times2\times 1$ ISD 
CDW phase in the CsV$_3$Sb$_5$ system are further characterized as higher-order-like VHSs, where a higher-order VHS is defined by a zero value of the Hessian determinant\,\cite{Yuan2019}, serving as sources of diverse intriguing many-body phenomena\,\cite{Yuan2019,classen2024high}.
We note that both the type-II and higher-order like VHSs found in our study are different from those studied in the pristine phase at the $k_z=0$ plane in the monolayer limit\,\cite{Kim_2023_ML_AVS} and in the bulk CsV$_3$Sb$_5$ system\,\cite{kang2022twofold}, respectively.
This suggests that the VHSs still play an important role in stacking-dependent 3D CDW phases.

We examine the $\pi$-phase shift stacking effect on the electronic structures of $A$V$_3$Sb$_5$.
We consider the $2\times2\times 2$ $\pi$-shifted ISD+ISD structure where alternating ISD layers along the $c$-direction are relatively shifted by a $\pi$-phase from each other [Fig.\,\ref{fig:structure}(g)]. The band dispersion and DOS for the $2\times2\times 2$ $\pi$-shifted ISD+ISD structure for all three alkali metals are shown in Fig.\,\ref{fig:ISDISD}. 
The same TB parameters used for the $2\times2\times 1$ non-$\pi$-shifted ISD structure yield TB bands in good agreement with the DFT bands of vanadium $d_{xy}$, $d_{3z^2-r^2}$, and $d_{x^2-y^2}$ orbitals  [Figs.\,\ref{fig:ISDISD}(a-c)]. 
To track the changes in energy bands under the $\pi$-phase shift, we compare the energy bands at $k_z = 0$ for the $2\times2\times2$ structure with those at both the $k_z = 0$ and $k_z = \pi$ planes for the $2\times2\times1$ structure. This comparison is necessary because, when a $\pi$-phase shift occurs, the energy bands at the $k_z = \pi$ plane in the $2\times2\times1$ CDW fold into the $k_z = 0$ plane in the $2\times2\times2$ CDW, resulting in a doubling of the number of energy bands.
The overall band dispersions at the \( k_z = 0 \) plane of the \( 2\times2\times2 \) ISD+ISD structure [Fig.\,\ref{fig:ISDISD}(a-c)] are similar to those of the \( 2\times2\times1 \) ISD structure at both the \( k_z = 0 \) and \( k_z = \pi \) planes [Figs.\,\ref{fig:ISDband}(a-c)]. However, a significant difference occurs in the flat-like dispersions along the \( A-L \) line in the \( 2\times2\times1 \) structure. In the \( 2\times2\times2 \) structure, these flat-like dispersions become more dispersive under the \( \pi \)-phase shift, as seen along the \( \Gamma-M \) line.
Consequently, the DOS in the $\pi$-shifted ISD+ISD structure exhibits a depletion near the Fermi level compared to the non-$\pi$-shifted ISD structure, indicating the lifting of the VHSs due to the $\pi$-phase shift [Figs.\,\ref{fig:ISDISD}(d-f)]. 
This leads to electrons occupying lower energy levels, thereby stabilizing the $\pi$-shifted structure.
The in-plane $d_{xy}$, $d_{3z^2-r^2}$, and $d_{x^2-y^2}$ orbitals capture essential changes in the DOS under the out-of-plane $\pi$-phase shift, highlighting their crucial roles in stabilizing the ground state 3D CDW order.

\begin{figure}[t!]
\centering
\includegraphics[width=.5\textwidth]{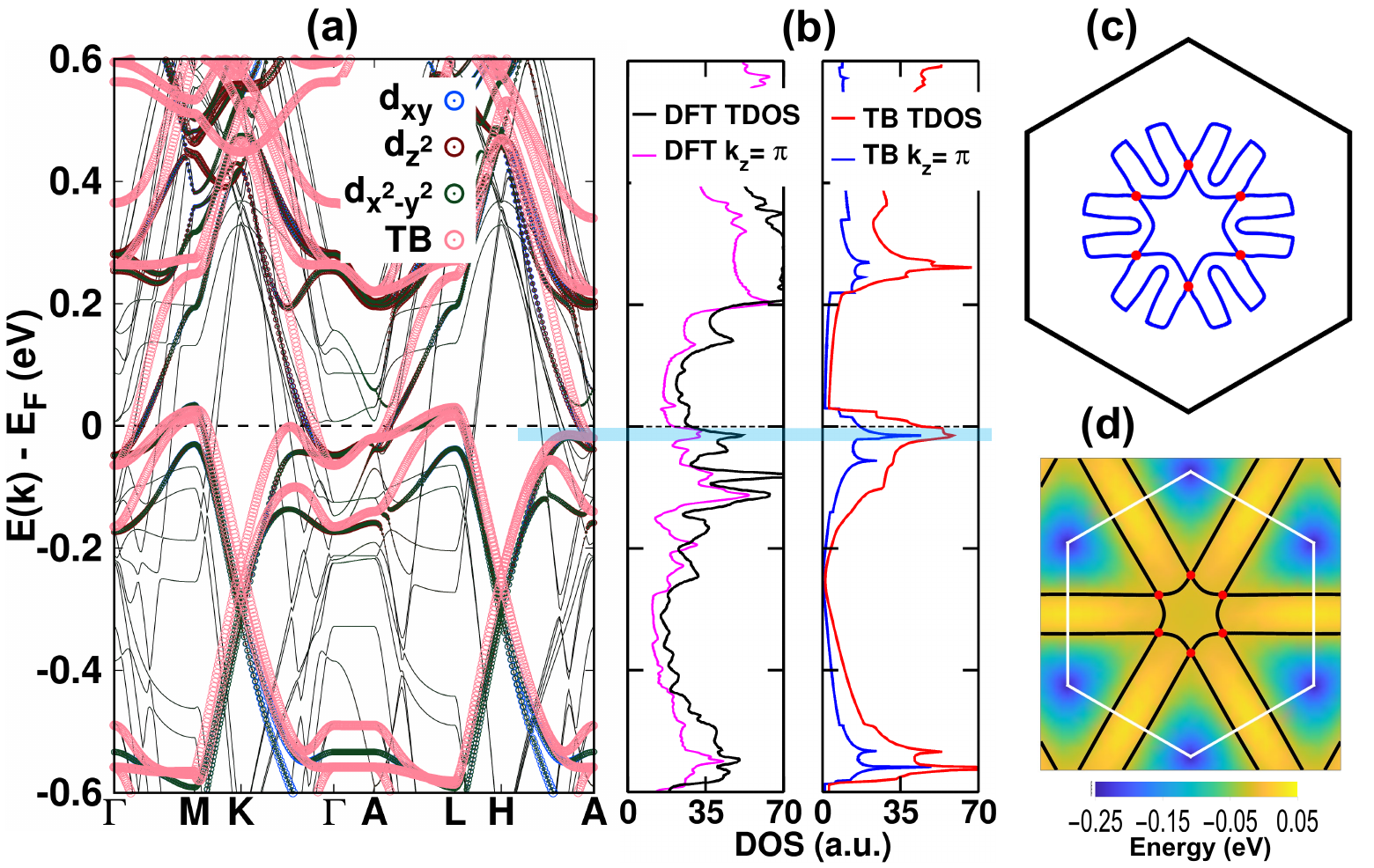}
\caption{Electronic structure of the $2 \times 2 \times 2$ SD+ISD structure without a $\pi$-phase shift for CsV$_3$Sb$_5$. (a) The orbital-projected DFT band structure, overlaid with the TB band structure.
(b) Total density of states (TDOS) and $k_z=\pi$ plane-resolved DOS from DFT and TB.
(c,d) Constant energy contours in the $k_z=\pi$ plane obtained using (c) DFT and (d) TB, respectively. Energy contours are displayed at energy values of -0.009 eV for TB and -0.012 eV for DFT calculations. Red colored dots designate the VHS points in the energy contours.
}
\label{fig:SDISD}
\end{figure}

We finalize our discussion by speculating on the significance of our findings in alternating SD and ISD distorted layers, which are observed as part of the $2\times2\times4$ CDW structure in CsV$_3$Sb$_5$ \cite{kang2023charge_six_breaking_7,ortiz2021fermi,hu2022coexistence}. 
The existence of the $2 \times 2 \times 4$ CDW structure in CsV$_3$Sb$_5$ was controversial, and was later found to be sensitive to growth conditions such as chemical disorder and thermal annealing\,\cite{stahl2022temperature, xiao2023coexistence, kautzsch2023structural}.
For the structural model, initial experiment suggested alternating ISD layers followed by three SD distorted layers along the $c$-direction without any $\pi$-phase shift between adjacent kagomé layers \cite{ortiz2021fermi}. However, subsequent studies \cite{xiao2023coexistence, kautzsch2023structural,hu2022coexistence} propose various stacking configurations, including complex stacking arrangements of ISD layers without alternating SD layers in the $2 \times 2 \times 4$ CDW structure. 
Our total energy calculations show that the $\pi$-shift stacking remains favorable over the SD+ISD stacking (Table\,\ref{table:sd+isd}), consistent with previous DFT calculations \cite{tan2021charge,xiao2023coexistence}.  
Our calculations of the band structure, DOS, and constant energy contour for CsV$_3$Sb$_5$ with alternating SD and ISD layers (Fig.\,\ref{fig:SDISD}), reveal a sharp DOS peak just below the Fermi level originating from the $k_z = \pi$ plane, attributed to the VHSs. This observation potentially supports the $2 \times 2 \times 4$ structural model with an ISD+ISD $\pi$-shift structure, as the electronic instability is indicated in the alternating SD+ISD structure and the additional periodicity doubling along the $c$-axis may relieve this diverging DOS at the Fermi level as in the case of the $2 \times 2 \times 1$ to $2 \times 2 \times 2$ transition.

\section{Conclusion}
Our comprehensive investigation, employing DFT and tight-binding calculations, delves into the stacking-dependent electronic structures of 3D CDW structures in $A$V$_3$Sb$_5$. 
Our findings unveil the electronic origin of out-of-plane CDW ordering, highlighting the critical role of interlayer coupling in electronic instabilities near the Fermi level.
A crucial aspect of our findings centers around the instability inherent in the $2\times 2\times 1$ ISD structure, attributed to the intense density of states originating from the $k_z = \pi$ plane at or near the Fermi level. This enhanced density of states is associated with the presence of type-II van Hove singularities. 
However, upon the 3D stacking of ISD layers along the $c$-direction with a $\pi$-phase shift, we observe a substantial renormalization of the density of states at the Fermi level, leading to the disappearance of the van Hove singularities and thus stabilizing the 3D $\pi$-shifted ISD+ISD CDW ground state.
Furthermore, our analysis highlights that the alternative stacking of ISD+SD layers in CsV$_3$Sb$_5$ also presents electronic instabilities near the Fermi level, making it energetically less favorable than the ISD+ISD $\pi$-shifted configuration.

Additionally, we have formulated an effective 3D tight-binding model that captures the essential physics of $A$V$_3$Sb$_5$ kagomé systems. This model serves as a valuable platform for further investigation of 3D interlayer coupling effects on novel phenomena observed in $A$V$_3$Sb$_5$ kagomé metals. We anticipate that this approach will uncover new exotic physics beyond the current studies limited to 2D kagomé models, such as three-dimensional chiral flux order and three-dimensional electronic nematicity.

\yk{We conclude by comparing our study to a previous one \cite{Jeong22p235145, Sim24p15} that introduces a comprehensive tight-binding model for the four \(d\) orbitals of V and the \(p_z\) orbital of Sb, which is essential for describing the energies of four VHSs near the Fermi level in the \(k_z=0\) plane under various strains and might be useful for describing all VHSs throughout the BZ. While this model is comprehensive, our simple model focuses on the \(d_{xy}\) orbitals and effectively describes the VHS that governs the low-energy electronic structure in the \(k_z=\pi\) plane without 3D CDW. We believe our simplified TB theory offers a practical approach for future research on interacting Hamiltonians and complex systems, such as twisted bilayer $A$V$_3$Sb$_5$ systems.
}

\section*{Data availability statement}
All data are included within the article and its supplementary materials.

\section*{Declaration of competing interest}
The authors declare that there are no competing financial interests or personal relationships that could have influenced the work reported in this paper.

\section*{Acknowledgments}
Y.K. acknowledges the support from the National Research Foundation (NRF) of Korea under grant number NRF-2021R1A2C1013871. The Korea Institute of Science and Technology Information (KISTI) (KSC-2021-CRE-0116) provided the computational resource. 

\section*{Appendix A. Supplementary data}
Supplementary data to this article can be found online at

\def\urlprefix{}
\def\href#1#2{#2}
\balance
\bibliographystyle{elsarticle-num} 

\ifarXiv
    \foreach \x in {1,...,\numbersupplementpages}
    {
        \clearpage
        \includepdf[pages={\x}]{\supplementfilename}
    }
\fi

\end{document}